\documentstyle[12pt
]{article}

\begin{document}


\let\sss=\l                     

\def\vu{\varepsilon}
\def\a{\alpha}
\def\b{\beta}
\def\c{\chi}
\def\d{\delta}
\def\e{\epsilon}                
\def\f{\phi}                    
\def\g{\gamma}
\def\h{\eta}
\def\i{\iota}
\def\j{\psi}
\def\k{\kappa}                  
\def\l{\lambda}
\def\m{\mu}
\def\n{\nu}
\def\o{\omega}
\def\p{\pi}                     
\def\q{\theta}                  
\def\r{\rho}                    
\def\s{\sigma}                  
\def\t{\tau}
\def\u{\upsilon}
\def\x{\xi}
\def\z{\zeta}
\def\D{\Delta}
\def\F{\Phi}
\def\G{\Gamma}
\def\J{\Psi}
\def\L{\Lambda}
\def\O{\Omega}
\def\P{\Pi}
\def\Q{\Theta}
\def\S{\Sigma}
\def\U{\Upsilon}
\def\X{\Xi}


\def\ca{{\cal A}}
\def\cb{{\cal B}}
\def\cc{{\cal C}}
\def\cd{{\cal D}}
\def\ce{{\cal E}}
\def\cf{{\cal F}}
\def\cg{{\cal G}}
\def\ch{{\cal H}}
\def\ci{{\cal I}}
\def\cj{{\cal J}}
\def\ck{{\cal K}}
\def\cl{{\cal L}}
\def\cm{{\cal M}}
\def\cn{{\cal N}}
\def\co{{\cal O}}
\def\cp{{\cal P}}
\def\cq{{\cal Q}}
\def\car{{\cal R}}
\def\cs{{\cal S}}
\def\ct{{\cal T}}
\def\cu{{\cal U}}
\def\cv{{\cal V}}
\def\cw{{\cal W}}
\def\cx{{\cal X}}
\def\cy{{\cal Y}}
\def\cz{{\cal Z}}

\def\qq{\qquad}
\def\th{\tilde h }
\def\cl{{\rm cl}}
\def\Z{{\bf Z}}
\def\Ahb{\hat{\!\bar A}}
\newcommand{\Zzbar}
  {\mathord{\!\setlength{\unitlength}{0.9em}
  \begin{picture}(0.6,0.7)
  \thinlines
  \put(0,0){\line(1,0){0.6}}
  \put(0,0.75){\line(1,0){0.575}}
  \multiput(0,0)(0.0125,0.025){30}{\rule{0.3pt}{0.3pt}}
  \multiput(0.2,0)(0.0125,0.025){30}{\rule{0.3pt}{0.3pt}}
  \put(0,0.75){\line(0,-1){0.15}}
  \put(0.015,0.75){\line(0,-1){0.1}}
  \put(0.03,0.75){\line(0,-1){0.075}}
  \put(0.045,0.75){\line(0,-1){0.05}}
  \put(0.05,0.75){\line(0,-1){0.025}}
  \put(0.6,0){\line(0,1){0.15}}
  \put(0.585,0){\line(0,1){0.1}}
  \put(0.57,0){\line(0,1){0.075}}
  \put(0.555,0){\line(0,1){0.05}}
  \put(0.55,0){\line(0,1){0.025}}
  \end{picture}}}

\newcommand{\extraspace}{\addtolength{\abovedisplayskip}{2mm}
                        \addtolength{\belowdisplayskip}{2mm}
                        \addtolength{\abovedisplayshortskip}{2mm}
                        \addtolength{\belowdisplayshortskip}{2mm}}
\newcommand{\be}{\begin{equation}\extraspace}
\newcommand{\bestar}{\[\extraspace}
\newcommand{\ee}{\end{equation}}
\newcommand{\eestar}{\]}
\newcommand{\bea}{\begin{eqnarray}\extraspace}
\newcommand{\beastar}{\begin{eqnarray*}\extraspace}
\newcommand{\eea}{\end{eqnarray}}
\newcommand{\eeastar}{\end{eqnarray*}}

\newcommand{\nonu}{\nonumber \\[2mm]}

\newcommand{\strutje}{\rule[-1.5mm]{0mm}{5mm}}
\newcommand{\dis}{\displaystyle}
\newcommand{\bz}{\bar{z}}
\newcommand{\de}{\delta}
\newcommand{\del}{\partial}
\newcommand{\delb}{\bar{\partial}}
\newcommand{\bdel}{\bar{\partial}}
\newcommand{\nab}{\overline{\nabla}}
\newcommand{\dz}{\partial_z}
\newcommand{\zw}{(z-w)}

\newcommand{\half}{{\textstyle\frac{1}{2}}}
\newcommand{\quart}{\frac{1}{4}}
\newcommand{\halfN}{\frac{N}{2}}

\newcommand{\ef}{{\rm eff}}
\newcommand{\ind}{{\rm ind}}

\newcommand{\edb}{\frac{1}{\bar\partial}}
\newcommand{\ddb}{\frac{\partial}{\bar\partial}}
\newcommand{\vgl}[1]{eq.(\ref{#1})}
\newcommand{\gv}{\gamma^5}
\newcommand{\gu}[1]{\gamma^{#1}}
\newcommand{\gd}[1]{\gamma_{#1}}
\newsavebox{\uuunit}
\sbox{\uuunit}
    {\setlength{\unitlength}{0.825em}
     \begin{picture}(0.6,0.7)
        \thinlines
        \put(0,0){\line(1,0){0.5}}
        \put(0.15,0){\line(0,1){0.7}}
        \put(0.35,0){\line(0,1){0.8}}
       \multiput(0.3,0.8)(-0.04,-0.02){12}{\rule{0.5pt}{0.5pt}}
     \end {picture}}
\newcommand {\unity}{\mathord{\!\usebox{\uuunit}}}
\newcommand{\dr}{\raise.3ex\hbox{$\stackrel{\leftarrow}{\partial }$}}
\newcommand{\dl}{\raise.3ex\hbox{$\stackrel{\rightarrow}{\partial}$}}

\newcommand{\np}{Nucl.\ Phys.\ }
\newcommand{\pr}{Phys.\ Rev.\ }
\newcommand{\cmp}{Comm.\ Math.\ Phys.\ }
\newcommand{\pl}{Phys.\ Lett.\ }

\begin{titlepage}
\begin{flushright}
\noindent June 1993 \hfill LBL-33777, UCB-PTH-93/08\\
$\strutje$ \hfill  KUL-TF-93/25\\
$\strutje$ \hfill   hep-th/9306135\\
\end{flushright}

\vfill
\begin{center}
{\large\bf Regularisation of non-local actions \\in two-dimensional field
theories. }\\
\vspace{10mm}
{\bf \centerline{Alexander Sevrin${}^a$, Ruud
Siebelink\footnote{Aspirant NFWO, Belgium}${}^b$ and
Walter Troost\footnote{Bevoegdverklaard Navorser NFWO, Belgium}${}^{b}$}}
 \vskip .3cm
{\baselineskip = 12pt
\centerline{\sl{a. Department of Physics}}
\centerline{\sl{University of California at Berkeley}}
\centerline{\sl{and}}
\centerline{\sl{Theoretical Physics Group}}
\centerline{\sl{Lawrence Berkeley Laboratory}}
\centerline{\sl{Berkeley, CA 94720, U.S.A.}}}
\vskip .2cm
{\baselineskip = 12pt
\centerline{\sl{b. Instituut voor Theoretische Fysica}}
\centerline{\sl{Universiteit Leuven}}
\centerline{\sl{Celestijnenlaan 200D, B-3001 Leuven, Belgium}}}
\end{center}
\vfill
\begin{center}
{\bf Abstract}
\end{center}
\begin{quote}
\small
Taking the induced action for gauge fields coupled to affine currents
as an example, we show how loop calculations in non-local two-dimensional
field theories can be regulated.
Our regularisation method for one loop is based on the method of Pauli and
Villars. We use it to calculate the renormalisation factors
for the corresponding effective actions, clearing up some discrepancies in
the literature. In particular, it will be shown explicitly that vector
gauge transformation invariance and Haar invariance of functional integral
measures impose different requirements, but they are related by a
counterterm (which is local in terms of group variables). For higher loops,
we use the method of covariant derivatives combined with Pauli-Villars to
argue that the one loop result remains unaltered.

\normalsize
\end{quote}
\end{titlepage}
\parindent 5truemm
\parskip 0truemm

\newpage

\pagestyle{plain}
\renewcommand{\thepage}{\roman{page}}
\setcounter{page}{2}
\mbox{ }

\vskip 1in

\begin{center}
{\bf Disclaimer}
\end{center}

\vskip .2in

\begin{scriptsize}
\begin{quotation}
This document was prepared as an account of work sponsored by the United
States Government.  Neither the United States Government nor any agency
thereof, nor The Regents of the University of California, nor any of their
employees, makes any warranty, express or implied, or assumes any legal
liability or responsibility for the accuracy, completeness, or usefulness
of any information, apparatus, product, or process disclosed, or represents
that its use would not infringe privately owned rights.  Reference herein
to any specific commercial products process, or service by its trade name,
trademark, manufacturer, or otherwise, does not necessarily constitute or
imply its endorsement, recommendation, or favoring by the United States
Government or any agency thereof, or The Regents of the University of
California.  The views and opinions of authors expressed herein do not
necessarily state or reflect those of the United States Government or any
agency thereof of The Regents of the University of California and shall
not be used for advertising or product endorsement purposes.
\end{quotation}
\end{scriptsize}

\vskip 2in

\begin{center}
\begin{small}
{\it Lawrence Berkeley Laboratory is an equal opportunity employer.}
\end{small}
\end{center}

\newpage
\renewcommand{\thepage}{\arabic{page}}
\setcounter{page}{1}

\baselineskip=14pt
\section{Introduction}
In two-dimensional field theory, the chiral algebra that constitutes (part
of) the symmetry algebra is of central interest.  Classically, coupling
gauge fields to that current algebra introduces a local invariance, which
in turn can be used to gauge away the gauge fields, so that these do not
acquire a dynamical existence.  Quantum mechanically however, due to
anomalies, these gauge fields often do become propagating fields,
describing  new degrees of freedom.  The best known case is of course the
Liouville mode for non-critical strings.  The action used as a starting
point to describe the quantum theory of these new degrees of freedom is the
induced action, i.e.\ the action resulting from integrating out the
`matter' degrees of freedom that make up the chiral currents.  In the case
of linear chiral algebras, where the central charge $c$ only appears in the
most singular term of the operator product expansions, this induced
action is proportional to
$c$. For non-linear algebras, the central charge also appears elsewhere in
coupling constants, and the induced action can be expanded in a power
series, $\Gamma _{\ind} = {\displaystyle \sum_{i =
0}^\infty}c^{1-i}\Gamma ^i$ where $\Gamma ^0$ will be called the
`classical' term.  Building a quantum theory on this induced action, the
following remarkable renormalisation property has been noticed: the quantum
effective action has the same functional form as the classical term
in the induced action, but a renormalisation of fields and couplings takes
place :
\begin{equation}
\Gamma _{\ef}[\Phi ] = Z_\Gamma \Gamma ^0[Z_\Phi \Phi ]\,.
\label{renorGamma}
\end{equation}
This property has been verified for Wess-Zumino-Witten models \cite{wzw} in
\cite{polouche}, for two-dimensional gravity
in \cite{kpz}, for $W_3$ gravity in \cite{ssvnc} to first order and to all
orders in \cite{dbg}. It was recently extended \cite{zfactors}, also to all
orders, to {\it all} extensions of two-dimensional
gravity that can be obtained from a Drinfeld-Sokolov reduction
 of WZW models based on (super)algebras \cite{drisok}.

The WZW models are the simplest examples, and in view of the Drinfeld-Sokolov
reduction technique, at the same time also generic. For these models,
there actually exist several different methods to obtain \vgl{renorGamma}.
We will recapitulate the different methods in the next section. These
methods agree as to the general form of the result, and also as to the
value of $Z_\Gamma $.  However, they disagree on the field renormalisation
factors $Z_\Phi $.  In view of the universal applicability of
\vgl{renorGamma},
valid as it seems to be for a very wide class of chiral algebras, this is
somewhat disconcerting.  As will become clear, each of these methods can be
criticized on the grounds that little attention is payed to one of the
prominent features of quantum field theory, namely
the need to make sense of the divergent expressions as they are encountered
 in perturbation theory. Partly this is due to the fact that
sometimes divergent diagrams, when treated in a cavalier fashion, formally
lead to reasonable finite expressions, which are then taken to be
the final answer. However, certainly since the advent of anomalies, we have
learned that even these finite parts may turn out to have a value that is
different from the 'naive' one.  Perhaps the simplest example is the box
diagram contribution to photon-photon scattering in QED: gauge invariance
requires it to vanish at zero momenta, but a simple four-dimensional
calculation leads to a finite integral that is non-zero. Needless to say
a regularised version (whether Pauli-Villars, dimensional regularisation,
or others) leads to the correct (zero) value.
It will be shown in this paper that a similar phenomenon occurs for the
(finite) renormalisation factors for induced two-dimensional gauge theories.

In order to resolve the discrepancy then, one has to face the issue of
regularisation in non-local field theories. Indeed, the induced actions are
typically non-local. In local field theory, one has developed by now a
whole arsenal of techniques to deal with the divergences that occur in a
perturbative expansion of Green functions (or $S$-matrix elements). They
typically involve the introduction of a cut-off at high momenta. Then
some renormalisation conditions are imposed, defining the physical
parameters of the theory in terms of the parameters in the Lagrangian as
well as the cut-off. Finally, the cut-off is taken to infinity keeping
the physical parameters equal to their experimental values.
Alternatively, one introduces the physical parameters in the Lagrangian
from the start, and adds counterterms to it, order by order in
perturbation theory, to keep these parameters at their physical values.
A crucial role is played in this whole development by the locality
properties: all counterterms added, finite or infinite (i.e. terms with or
without limit when the cut-off is removed), are local expressions in terms
of the fields. In non-local field theories one loses this important criterium.
We will take a very pragmatic attitude here, and first look for a method
that regularises the theory. Rather than investigating all possible
 counterterms that one may reasonably (want to) add, we will at first not
add any when none is needed.
In this paper we will also limit ourselves to  very specific
non-local field theory,
{\it viz.} the two-dimensional actions, induced by matter, of chiral gauge
fields that couple to matter currents forming an affine Lie algebra. After
introducing the model in section two, we will present three derivations of
the renormalisation formula \vgl{renorGamma} in section three. In the next
section, based on simple considerations of the diagrams to be computed for
the two point function, we will introduce a regularisation for the quantum
theory,
including single loops only. This regularisation can be translated to a
Pauli-Villars style regularisation \cite{Pauli}, which will have the
unusual property
that it is based on a non-local mass term. Then we will go on to show that
it regularises {\it all} one-loop diagrams, and we will compute the
$n$-point functions in the same 1-loop approximation. In section six we
will compare
the resulting renormalisation factors with the formal calculations of the
third section. Also in that section we will relate our computation
to computations in the WZW model, which is closer to a local field theory
than the induced actions for gauge fields we consider up to there. We will
show
that the regularisation we adopted is fully compatible with the formal
Haar-invariance of the WZW integration measure, provided one includes a
{\it finite  local} counterterm.
In section seven we propose a regularisation method that goes beyond one
loop,
introducing a method based on higher derivative terms in the Lagrangian
(which are introduced covariantly), followed by a Pauli-Villars style
regularisation of the remaining one-loop divergences. We will show that the
one-loop result obtained in section five remains unaltered, and
that it in fact provides the full answer.

\section{Induced Wess-Zumino-Witten models}

The induced action $\Gamma[A]$ for a Lie algebra valued gauge field $A$
is defined to be the
generating functional of all connected current-current correlators for a
given holomorphic affine Lie algebra of level $k$:
\footnote{We use the following conventions: If $[t_a,t_b] =
f_{ab}^{\;\;\;c}\,t_c$, then $f_{ac}^{\;\;\;d} f_{bd}^{\;\;\;c}=-\tilde h\,
g_{ab}$, where $\tilde h$ is the dual Coxeter number. In a
representation $R$ we have $\mbox{tr} \{ t_at_b\}=-x\,g_{ab}$, with $x$
the index of the representation}
\be
e^{\dis -\G[A]} = \, \langle
  e^{\dis - \frac{1}{\p x} \int d^2 z \;
 \mbox {tr} \left\{ J(z) A(z) \right\}} \rangle , \label{defgam}
\ee
and correspondingly, for an anti-holomorphic algebra one defines the
induced action $\bar\Gamma[\bar A]$ for a gauge field of  opposite
chirality.
These induced actions arise when integrating out 'matter' degrees of
freedom, indicated by the $\langle \rangle$ signs. All that is needed
from this underlying model are the operator product expansions for the
currents $J^a$, which we take to form an affine Lie algebra. Explicitly,
\begin{eqnarray}
\Gamma[A] &=& \frac{k}{2\pi x} \int d^2z \,\, \mbox{tr} \left\{ \,\,
\frac{1}{2} \, A \, \ddb A \, - \frac{1}{3} \, \edb A \, \left[ A, \ddb A
\right] \right. \nonu
&& \left. - \, \frac{1}{4} \,  \edb A \, \left[ A, \edb \left(
\left[ A, \ddb A \right] \right) \right] + \cdots \right\} .\label{indact}
\end{eqnarray}
Although this action is non-local, its variation under the gauge
transformation
\be
\d_{\h } A =\bar\partial \h - [A, \h]
      = \bar D[A]\eta \ ,
\label{varA}
\ee
is a purely local expression, given by
\be
\d_{\h } \G [A] = - \frac{k}{2\p x} \int d^2 z \, \mbox{tr} \left\{ \h
\del A \right\}.
\label{vargam}
\ee
This transformation property in fact contains all the information
needed concerning the induced action.
Introducing the "induced current" $u$
\be
u_a(x)= - 2\p \frac{\d \G[A]}{\d A^a(x)},
\label{defu}
\nonumber
\ee
one finds from eq.\ (\ref{vargam}) the Ward identity
\be
\bar D[A] \, u/k = \del A .
\label{ward}
\ee
In \cite{polwi} it was realised that this Ward identity is nothing
but the statement that the curvature for the Yang-Mills field
with components $\{A,u/k\}$ vanishes. This condition is then solved by
parametrizing $A\equiv \bar{\del} g g^{-1}$ and $u/k\equiv\del g g^{-1}$.
Rewriting the righthand side of eq.\ (\ref{vargam}) using $A =
\bar{\del} g g^{-1}$ and $\h =\d_{\h }g g^{-1}$ one obtains
the equation of motion of a Wess-Zumino-Witten model. One concludes that
\be
\G [A=\bdel g g^{-1}]=-k\ S^+[g]
\nonumber
\ee
with
\be
S^+ [g]=  \frac{1}{4\p x} \int d^2 z \;
\mbox{tr} \left\{ \del g^{-1} \bar{\del} g \right\}
+ \frac{1}{12\p x} \int d^3 x\; \e^{\a\b\g} \,
\mbox{tr} \left\{ g_{,\a} g^{-1} g_{,\b}
g^{-1} g_{,\g} g^{-1} \right\}.
\label{splus}
\ee
\newline
Having established a dynamical, though non-local, action for the gauge
field $A$, one proceeds with the subsequent quantisation
of this theory. The objects of interest are the generating functionals
$Z[u]$ and $W[u]$ of the Green functions of this quantum theory:
\be
Z[u]=e^{\dis -W[u]}=\int \cd A \, e^{\dis -\G[A]+\frac{1}{2\p x}
  \int  \; \mbox{tr} \left\{ u\, A \right\}  } , \label{defw}
\ee
or, equivalently, the Legendre transform $\Gamma $:
\be
\Gamma [A_\cl]=\min_{\{ u\} }\left( W[u]-\frac{1}{2\pi x}\int
\mbox{tr} \{ u\,A_{\cl}\} \right).
\ee
which is the generating functional for one particle irreducible
diagrams, and is called the effective action. For brevity, we will use
the same name for $W[u]$ also.
In the sequel we will also use the level-independent reference functional
$ \G^0[A]=\G[A]/k $ and its
Legendre transform
\be
W^0[u]=\min_{\{ A\} }\left( \G^0[A]-\frac{1}{2\pi x}\int
\mbox{tr} \{ u\,A\} \right).
\label{defwnul}
\ee
The explicit form of this last quantity is given by
\be
W^0[u\equiv \del g g^{-1}]=S^+[g^{-1}].\label{qquan1}
\ee

The opposite chirality can be treated along the same lines. We have that
$\bar\G^0[\bar A= \del \bar g \bar g^{-1}]=-S^-[\bar g]$. Because of the
identity $S^-[\bar g]=S^+[\bar g^{-1}]$, we have also that
\be
W^0[\bar A ]=-\bar\G^0[\bar A ]. \label{WisGbar}
\ee

\section{Three Derivations of the Effective Action}

In this section we review three different
methods to calculate the effective action $W[u]$.

The first method is a semiclassical approximation \cite{zamo2,stony}.
According to the steepest descent
method the effective action \vgl{defw} is approximated by
\begin{equation}
e^{\dis -W[u]}
 \simeq  e^{\dis -\G[A_{\rm cl}]+\frac{1}{2\p x}
  \int  \; \mbox{tr} \left\{ u\, A_{\rm cl} \right\}}
 \int {\cal D} \a \,\,e^{\dis - \half \int \a^a \frac{\d^2 \, \Gamma^{\,0}
 [A_{\rm cl}]} {\d A_{\rm cl}^{\,a} \, \d A_{\rm cl}^{\,b}} \, \a^b } \,,
\label{effact}
\end{equation}
where $A_{\rm cl}[u]$ is the saddle point and  $\a$ is the fluctuation
around this point. The position of the saddle is obtained by solving
\begin{equation}
-2\p\frac{\d \Gamma^{\,0} [A_{\rm cl}]}{\delta A_{\rm cl}^{\,a}} =
\frac{u_{\,a}}{k}\,, \label{u in A}
\end{equation}
giving $A_{\cl}$ as a functional of $u$.
  Then, all that
has to be done is to compute the functional determinant:
\begin{equation}
W[u] \simeq kW^0[u/k] + \half \log \det
  \frac{\delta^2\Gamma [A_{\rm cl}]}{\delta A_{\rm cl}\delta A_{\rm cl}}\,.
  \label{Wsemicl}
\end{equation}
To evaluate this expression we take advantage
of the Ward identity. Indeed, taking the functional derivative of eq.\
(\ref{ward}) with respect to $A_{\rm cl}$, one obtains the
operator identity
\begin{eqnarray}
-2\pi \bar D_x[A_{\rm cl}]^{ac} \frac{\d^2 \, \Gamma^{\,0} [A_{\rm cl}]}
{\d A_{\rm cl}^{\,c}(x) \, \d A_{\rm cl}^{\,b}(y)} =
D_x[u/k]^a_{.\,\,b} \,\delta^2(x-y) \,,
\label{operator}
\end{eqnarray}
where $D[u/k]$ is the covariant derivative $\partial - u/k$, and the
subscript $x$ denotes the coordinate the differential operator is
acting on.
It follows that the sought-after determinant is equal to the ratio
of two determinants involving only ordinary covariant derivatives:
\begin{eqnarray}
\mbox{det} \frac{\d^2 \, \Gamma^{\,0} [A_{\rm cl}]}
{\d A_{\rm cl}\, \d A_{\rm cl}} =
\frac{\mbox{det}D[u/k]}{\mbox{det}\bar D[A_{\rm cl}]}
\label{quotient}
\end{eqnarray}
These determinants can then be computed \cite{orlando} by writing them out
as pathintegrals over $bc$ ghost-antighost systems. The
ghostsystems realise an (anti-) holomorphic current
algebra with level $k_{ghost}= 2\tilde h$. Therefore,
\begin{eqnarray}
\log\det D[u/k]&=& 2\tilde h W^{\,0} [u/k] \nonu
\log\det \bar D[A_{\cl}]&=& -2\tilde h \Gamma^{\,0}[A_{\rm cl}] \nonu
\half\log\det \frac{\d^2 \, \Gamma^{\,0} [A_{\rm cl}]}
{\d A_{\rm cl} \, \d A_{\rm cl}}
&=& 2\tilde h W^{\,0}[u/k] - \frac{\tilde h}{k} \int u_{\,a}
\frac{\d \, W^{\,0} [u/k]} {\d u_{\,a}/k} \,.
\label{detexplicit}
\end{eqnarray}
This gives the semiclassical result, valid through order $k^0$:
\begin{eqnarray}
W[u] = (k+2\tilde h) W^0\left[ \frac{u}{k+\tilde h}\right] \,.
\label{scresult}
\end{eqnarray}

The second method is called the KPZ \cite{kpz} approach (we adapt
the treatment in \cite{stony}). One starts from the following action,
containing gauge fields of both chiralities,
which is invariant under the transformation of $A$ as in eq. (\ref{varA})
 and $ \d_{\h}\bar A=D[\bar
A]\h $:
\be
\G [A]+\bar{\G}[\bar A]-\frac{k}{2\pi x}\int
\mbox{tr}\left\{A\,\bar A\right\} \,.
\label{covindaction}
\ee
The invariance follows from \vgl{vargam}.
This action is taken as a starting point for quantisation.
Because of the invariance, gauge fixing is necessary. The gauge freedom
is used to put the gauge field $\bar A$ equal to some fixed but arbitrary
value $\Ahb$. Introducing the corresponding ghosts,
one obtains the BRST-invariant action:
\be
\G[A,\Ahb,b,c]=\G [A]+\bar{\G}[\Ahb]-
\frac{k}{2\pi x}\int \mbox{tr}\left\{\Ahb \,A\right\}
+\frac{1}{2\pi}\int  bD[\Ahb] c \,.\label{BRSinvact}
\ee
We consider the partition function corresponding to this action,
\begin{equation}
\tilde Z[\Ahb] =\int \cd A\cd b  \cd c \; e^{\dis - \G[A,\Ahb,b,c] }.
\label{defZ}
\end{equation}
The partition function is independent of the gauge choice - in this case
this means independent of the function $\Ahb$ - and can be
normalised
to one. Expressing  the term $\bar{\G}[\Ahb]$ of the gauge
fixed action as
a pathintegral over a right-handed fermionic mattersystem with level $k$
one obtains
\begin{eqnarray}
1=\tilde Z[\Ahb]
&=&\int \cd
A\; e^{\dis
- \G[A] +\frac{k}{2\pi x}\int \mbox{tr}\left\{\Ahb \,A\right\} }
\nonumber\\
&& \;\;\;\int \cd b  \cd c \cd\bar\psi\; e^{\dis -\frac{1}{2\p
} \int bD[\Ahb] c + \bar\psi^t D[\Ahb] \bar\psi} .
\label{Zis1}
\end{eqnarray}
The background field $\Ahb$ couples to a current that consists of
three parts: matter ($\bar\j$), ghost ($bc$) and
gauge system ($A$) currents. The first two are antiholomorphic affine
currents with central
extensions $k$ and $2\tilde h$ respectively. The $\Ahb$-independence then
implies that the gauge part, $J_{\rm gauge}=-\frac{k}{2}A$ also constitutes
an anti-holomorphic affine algebra with a compensating value of the
central charge, {\it i.e.} with level $-(k+2\tilde h)$.
Therefore, combining \vgl{defw}, \vgl{Zis1} and the anti-holomorpic
counterpart of \vgl{defgam} we obtain
\begin{eqnarray}
e^{\dis -W[u=k\Ahb]} &=&
\int \cd A e^{\dis
- \G[A] +\frac{k}{2\pi x}\int \mbox{tr}\left\{\Ahb \,A\right\} }
\nonumber\\
&=& \langle \exp \, - \frac{1}{\p x} \int d^2 x \;
 \mbox {tr} \left\{\Ahb \,J_{\rm gauge} \right\} \rangle \nonumber
\\ &=& e^{\dis -(-k-2\tilde h)\bar \G^0[\Ahb]},
\label{reskpz}
\end{eqnarray}
and using \vgl{WisGbar},
\be
W[u] = (k+2\tilde h) W^0\left[ \frac{u}{k}\right] \,.
\label{kpzresult}
\ee
This ends the second derivation.

\vspace{2mm}
The third method \cite{polouche} consists first of all in a change of
variable from the gauge field $A$ to group variables $g$,
\be A=\bar{\del} g g^{-1}. \label{Anaarg} \ee
{}From $ \d_{\h} A = \bar D[A]\h = \bar D[A] \left( \d_{\h} g g^{-1}\right) $
one infers that the following jacobian is picked up :
\begin{eqnarray}
\cd A &=& \cd g\det \bar{D}[A=\bdel g g^{-1}]\nonumber\\
      &=& \cd g \; e^{\dis 2\tilde{h}S^+[g]} \,.
\label{maat}
\end{eqnarray}
Then from \vgl{defw},
\begin{eqnarray*}
e^{\dis -W[u]}=\int \cd g \; e^{\dis (k+2\tilde{h})S^+[g] +\frac{1}{2\p
x} \int  \; \mbox{tr} \{ u\bdel g g^{-1} \} }.
\label{pathint}
\end{eqnarray*}
If one parametrises the source $u\equiv (k+2\tilde h) \partial h h^{-1}$
and one uses the Polyakov-Wiegmann identity \cite{polwi},
\be
S^+[h^{-1}g]=S^+[h^{-1}]+S^+[g]+\frac{1}{2\pi x}\int \mbox{tr}\,\{
\del h h^{-1} \bdel g g^{-1} \} ,\label{pw}
\ee
one finds
\begin{eqnarray*}
e^{\dis -W[u]}=e^{\dis -(k+2\tilde{h})S^+[h^{-1}]}
\int \cd g \; e^{\dis (\k+2\tilde{h})S^+[h^{-1}g]} \,.
\end{eqnarray*}
Using now the invariance of the Haar measure,
this last pathintegral evaluates to a constant, so
\begin{eqnarray}
W[u] &=& (k+2\tilde h) W^0\left[ \frac{u}{k+2\tilde h}\right] \,.
\label{reshaar}
\end{eqnarray}

Let us take stock. We have presented three methods to get information on
the effective action, and obtained
eqs.(\ref{scresult},\ref{kpzresult},\ref{reshaar}).
Three times we were led to the same functional dependence, {\it viz.} that
the quantum effective action is a rescaling of the classical action. Three
times also, the resulting renormalisation factor for the action, which can
be seen to be a coupling constant renormalisation, has the same value,
$k \rightarrow k+2 \tilde h$. However, the field renormalisation factor
$Z_u^{-1}$ was three times different, $k \rightarrow k+n \tilde h$ with
$n=1,0,2$ respectively. It is clear that, to solve this puzzle, more
care is needed in the heuristic steps of the above methods. For that
purpose, we will introduce in the next section a scheme that allows us
to properly take into account the quantum field theory divergences
implicitly present in the above derivations.

The coupling constant renormalisation factor, common to the three methods,
is expected to be valid beyond the one loop approximation. The argument
rests on the multiple valuedness of the WZW functional $S^+$. To have in
eq.(\ref{defw}) a
univalued $Z[u]$-functional, the prefactor of $W^0$ has to be an integer.
Since higher loop quantum corrections will modify it with terms of order
$1/k$, they should be absent. Note that no similar argument can be given
for the field renormalisation factors. Thus both the fact that they
are different, and that they are closely related, is unexplained.

\section{Regularisation}

To investigate the quantum theory based on the induced action, one can
follow the canonical methods used for non-local actions. One
rewrites the action as a sum of monomials in the fields, \vgl{indact}. The
quadratic term determines the propagator, the higher order monomials the
vertices. The non-locality manifests itself only in the fact that the
propagator has an unfamiliar form, and the vertices contain inverse powers
of momenta. To decide on a regularisation  of the resulting diagrams is a
different matter however. One method would be to try and convert the
theory to a local one first. For the present case, this would involve the
change of variables as in \vgl{Anaarg}, which is possibly a sensitive
issue to start with. A more direct confrontation with the difficulties
was made in  \cite{loopsnlft}, in the framework of two-dimensional gravity
theories. There, each diagram was regularised, by hand, by inserting an
exponential cut-off in the momentum integration. This procedure is not
without difficulties, as discussed in \cite{loopsnlft}, with dependence of
the diagrams on routings of momenta.

The method that we are about to introduce will stick closely to traditional
methods based on the ideas of Pauli and Villars. We will at first have a
look at the simplest diagrams to be regulated. It will be shown that this
can be done straightforwardly by a compensating diagram with a modified
propagator. Then we determine which Lagrangian corresponds to this
modification. We regard this as a crucial step to ensure consistency of our
treatment. The regularisation of the remaining one-loop diagrams is
then completely fixed.

To see what we need, we start by considering the two-point function.
The momentumspace Feynman rules can be read off from
\begin{eqnarray*}
&-&(2\pi)^3 \,\frac{\d^2 \, \Gamma^{\,0} [A_{\rm cl}]}
{\d A_{\rm cl}^{\,a}(q) \, \d A_{\rm cl}^{\,b}(r)}
\equiv \Delta_{ab}(q,r) + \sum_{n=3}^{\infty }
V^n_{ab}[A_{\rm cl}](q,r) \\
&&\qq\qq= g_{\,ab} \,\frac{\bar q }{q}
\,\delta^2(q+r) \\
&&\qq\qq+ \frac{2i}{(2\pi)^2} \int d^2p \,\,f_{abc}
\,A_{\rm cl}^{\,c}(p) \,\,\delta^2(p+q+r) \,\,{\cal V}_3(q,r,p) \\
&&\qq\qq+ \frac{1}{(2\pi)^4} \int d^2p \,\,d^2s  A_{\rm cl}^{\,d}(p)
\,A_{\rm cl}^{\,e}(s) \,\, \delta^2(p+q+r+s) \\
&&\qq\qq\qq\qq
\left\{ \,f_{ad}^{\,\,\,\,\,c}\,f_{ceb}\,\,\,{\cal V}_4(q,r,p,s)
+ \,f_{ab}^{\,\,\,\,\,c}\,f_{cde}\,\,\,{\cal W}_4(q,r,p,s)\,\right\} \\
&&\qq\qq+ ...
\end{eqnarray*}
with
\begin{eqnarray*}
&&{\cal V}_3(q,r,p) = \frac{p\bar q- \bar p q}{pqr} \\
&&{\cal V}_4(q,r,p,s) =  \frac{-1}{p+q} \left\{
\frac{\,\,\,\bar r}{q \, r}- \frac{\,\,\,\bar r}{p \, r}+
\frac{\,\,\,\bar s}{p \, s}- \frac{\,\,\,\bar s}{q \, s}+
\frac{\,\,\,\bar p}{r \, p}- \frac{\,\,\,\bar p}{s \, p}+
\frac{\,\,\,\bar q}{s \, q}- \frac{\,\,\,\bar q}{r \, q} \right\} \\
&&{\cal W}_4(q,r,p,s) = \frac{-1}{p+s} \left\{
\frac{\,\,\,\bar r}{p \, r}- \frac{\,\,\,\bar s}{q \, s}+
\frac{\,\,\,\bar s}{r \, s}- \frac{\,\,\,\bar q}{p \, q} \right\} \\
\end{eqnarray*}

The two-point function at one loop order consists of the two diagrams of
figs.1 and 2.

\begin{figure}
\setlength{\unitlength}{1cm}
{\begin{picture}(12.7,4)(-1,0)
\put(4.55,2){\line(1,0){1.5}}
\put(6.75,2){\circle{6}}
\put(7.45,2){\line(1,0){1.5}}
\end{picture}}
\caption{Vacuum Polarisation Diagram}
\label{fig. 1}
\end{figure}
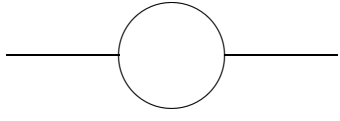

The (unregularised) expression for the vacuum polarisation diagram (fig.1)
is given by
\begin{eqnarray*}
&&-\frac{1}{4}\mbox{tr} \left\{ \Delta^{-1} V^3 \Delta^{-1} V^3\right\} \\
&&=\frac{-\tilde h}{(2\pi)^4} \int d^2p \int d^2q \,
  A_{\rm cl}^{\,a}(p) A_{{\rm cl}\,a }(-p) \frac{q^2}{|q|^2}
  \frac{(p+q)^2}{|p+q|^2} \frac{(\bar pq-\bar qp)^2}{q^2 (p+q)^2 p^2} \,,
\label{unregul}
\end{eqnarray*}
which is logarithmically divergent. We refrain from making formal
simplications before the integral is well-defined. To regularise it we
add a supplementary
diagram, with an identical integrand except for the replacement of the
propagator denominators
$|q|^2$ and $|p+q|^2$ by $|q|^2+M^2$ and $|p+q|^2+M^2$, and a
change in overall sign. This improves the convergence by two powers which
is sufficient in this case. We therefore adopt the following convergent
expression for this diagram
\begin{eqnarray}
\sum_{i=0} c_i \frac{-\tilde h}{(2\pi)^4} \int d^2p \int d^2q\,
  A_{\rm cl}^{\,a}(p) A_{{\rm cl}\,a }(-p) \frac{(\bar pq-\bar qp)^2}
  {p^2\,(|q|^2+M_i^2)(|p+q|^2+M_i^2)} \,,
\label{V3V3}
\end{eqnarray}
where the sum runs over the original $i=0$ term (i.e. $c_0=1$
and $M_0=0$), and over (all) the Pauli-Villars term(s) needed to
render the Feynman diagram well defined. We take
$\sum_{i=0} c_i=0$. After the momentum integration, the limit $M_i
\rightarrow \infty \, (i \ne 0)$ still has to be taken.

The Lagrangian
interpretation for these regularising steps is quite simple. The extra
contributions correspond to extra (PV) fields, having the same couplings
as the original ones but different propagators, and an extra factor $c_i$
for the loop. This  factor can be added by hand, or, perhaps more
attractively,
by limiting oneself to $c_i=\pm 1$ and considering $-1$ to result from
the sum of two fermionic ($-2$) and one bosonic ($+1$)
PV-field. The change in the propagator can be viewed as arising from a
massterm for the PV-fields $\a_{ia}$
\begin{equation}
\frac{-M_i^2 }{16\pi}\int \alpha_i^{\,a} \,\frac{1}{\bar\partial^2} \,
\alpha_{i\,a}
\label{mass}
\end{equation}
A non-local mass term has been used in a different context in
\cite{FrankRuud1}.
If one substitutes $A=\bar\del\varphi$ in \vgl{indact},
the quadratic terms (and the trilinear terms) become local in $\varphi$,
and the proposed mass term
has a completely conventional form. We adopt it from now on.

\begin{figure}
\setlength{\unitlength}{1cm}
{\begin{picture}(12.7,4)(-1,0)
\put(6.75,2.7){\line(1,1){1.1}}
\put(6.75,2){\circle{6}}
\put(6.75,2.7){\line(-1,1){1.1}}
\end{picture}}
\caption{Seagull Diagram}
\label{fig. 2}
\end{figure}
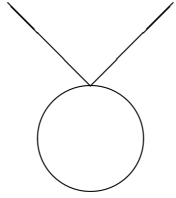

The expression for the seagulldiagram (fig.2) is now completely
fixed. One finds that
\begin{eqnarray}
&&\frac{1}{2}\sum_{i=0} c_i \mbox{tr} \left\{ \Delta_i^{-1} V^4 \right\}
\label{V4} \\
&&= \sum_{i=0} c_i \frac{-\tilde h}{(2\pi)^4} \int d^2p \,\int
d^2q\, A_{\rm cl}^{\,a}(p) A_{{\rm cl}\,a }(-p)
\frac{(\bar p+\bar q)(p-q)(\bar pq-\bar qp)}{p^2(|q|^2 + M_i^2)|p+q|^2 }
\,. \nonumber
\end{eqnarray}
Two remarks are in order. First, the linear divergence is again cured by the
condition $\sum_{i=0} c_i=0$. Second, note that the vertex denominators
are not altered, but only the propagators (for the PV contributions).

Both expressions (\ref{V3V3}) and (\ref{V4}) are thus made finite. It is
convenient to combine them before taking the limit $M_i\rightarrow
\infty$. This leads, after some elementary
algebra, to the following formula for the one loop effective
action:
\be
W_{\rm 1\,loop}[u(A_{\rm cl})] = \frac{-\tilde h}{(2\pi )^4}\int d^2p\,\,
A_{\rm cl}^{\,a}(p) \left(\, I_4 + I_{33} \,\right) A_{{\rm cl}\,a}(-p) +
\co(A_{\rm cl}^3)
\nonumber
\ee
with
\begin{eqnarray}
I_4 + I_{33}   &=& \sum_{i=0} \,c_i \int d^2q\,\frac{(\bar pq-\bar qp)^2}
{p^2(|q|^2+M_i^2)}\,\left\{ \frac{1}{|p+q|^2+M_i^2}
-\frac{1}{|p+q|^2} \right\}
\label{tweepunt}
\end{eqnarray}
We want to stress here that the crucial difference in the two
last denominators, is a direct consequence of the fact that the
contributions from  regulating fields are derived from a specific
Lagrangian.  As a result only {\it
propagators} are modified in the PV-method, whereas vertices
are simply duplicated, thus ensuring that
{\it all} the one-loop diagrams of the original theory are regularised. The
denominator coming from the non-local $V^4$-vertex therefore remains
"massless".
Formula (\ref{tweepunt}) also indicates that the ad-hoc
regularisation device of
multiplying the integrand with a standard convergence factor, would
give a vanishing result here.

To proceed with the evaluation of eq.(\ref{tweepunt}) we combine the
denominators using a Feynman parameter, shift the momenta and perform the
momentum integral to find
\begin{eqnarray}
\left.I_4 + I_{33}\right|_{M_i \rightarrow
\infty }&=& \sum_i c_i \frac{-2\pi |p|^2}{p^2} \int_0^1 d\alpha  \ln
\frac{\alpha M_i^2 + \alpha (1 - \alpha )|p|^2}{M_i^2 + \alpha (1 - \alpha
)|p|^2 }\nonumber\\
&=& \frac{-2\pi \bar p}{p}
\label{result1}
\end{eqnarray}
upon taking the limit.
This limit might perhaps have contained divergent
pieces (the individual contributions of the diagrams do). Then one would
have had to adjust the action with some ($M_i$-dependent)
counterterms, and include their contribution before taking the limit. These
counterterms are the ones that, in the usual renormalisation program,
renormalise the physical vs. bare parameters
in the Lagrangian. In the cases studied here, such divergent terms do not
appear. Therefore, we don't add any counterterms at present, and the
infinite mass limit gives directly the final answer:
\be
W_{\rm 1\,loop}[u(A_{\rm cl})] = \frac{-2\tilde h}{2\pi x} \int d^2z \,\,
\mbox{tr} \left\{ \,\, \frac{1}{2} \, A_{\rm cl} \, \ddb A_{\rm cl} \, +
\cdots \right\} \label{quad1loop}
\ee

\section{One Loop Calculation of the Effective Action}

Having established our regularisation method by looking at the two-point
function, we now compute the full effective action in the one loop
approximation. In principle, this could be done with diagrams also, but
we apply a simpler method. Let us consider the change in $W$ when we
perform a gauge transformation on its argument,
$\delta (u/k) =  D[u/k] \eta$. Since $A_{\cl}$ is
expressed in terms of it by \vgl{u in A}, this implies
$\delta A_{\rm cl} = \bar D[A_{\rm cl}] \eta$. If we make a simultaneous
transformation on the integration variables,
$\delta \,\alpha_i^{\,a} = f^a_{\,\,\,\,\,bc}\eta^b \alpha_i^{\,c}$, one
finds that
\begin{eqnarray*}
\delta \,\half \int \alpha_i^{\,a}
\,\frac{\partial^2 \, \Gamma^{\,0} [A_{cl}]}
{\partial A_{cl}^{\,a} \, \partial A_{cl}^{\,b}} \alpha_i^{\,b}
&=& 0.
\end{eqnarray*}
This can be seen most easily by combining the transformations in
$A=A_{\rm cl}+\a_i$ and using
\begin{eqnarray*}
\delta \left( \G^0[A] -\frac{1}{2\pi x}\int \mbox{tr}\left\{A\,
u/k \right\}\right) &=& \frac{1}{2\pi x} \int \mbox{tr}\left\{\eta\,
\bar\partial (u/k) \right\}.
\end{eqnarray*}
The gauge variation of $W[u]$ is then obtained from an anomaly calculation.
In the present setup, where jacobians for simultaneous transformation
of original and PV-fields cancel, this means that it is given by the
expectation value
of the variation of the PV-massterm (which is {\it not} invariant). We
follow the method of \cite{diaz}. Introducing a $T$-operator, by writing
the mass term \vgl{mass} as
\be
\half M_i^2 \int \alpha_i^{\,a} T_{a\,b} \, \alpha_i^{\,b}
\ee
one finds
\begin{eqnarray*}
\delta W_{\rm 1\,loop}[u(A_{\rm cl})] = \sum_{i=1} c_i
\mbox{Tr} \left\{ \eta \,\left( 1-T^{-1}\Gamma^{0''}/M_i^2 \right)^{-1}
\right\}\, ,
\end{eqnarray*}
where Tr denotes a matrix trace (which we have explicitly taken to be in
the adjoint representation) and a coordinatespace integration also.

To proceed, we recast the operator
$T^{-1}\Gamma^{0''}$ in a more convenient form, using covariant
derivatives and the operator identity \vgl{operator}. We find
\begin{eqnarray*}
T^{-1}\Gamma^{0''}&=& 4\left(\bar D + A_{\rm cl} \right)^2 \frac{D}{\bar D}
\,, \end{eqnarray*}
where $D=D[u/k]$ and $\bar D=\bar D[A_{\cl}]$.
It is important to realise that $D$ and $\bar D$ commute, due to
the zero curvature condition (\ref{ward}).
Therefore, they can be diagonalised
simultaneously. To evaluate the trace, one can make use of this property by
not using plane waves, but instead a basis of their eigenstates
(i.e. $g\,e^{ikx}$ when  $A_{\rm cl}=\bar\partial g\,g^{-1}$):
\begin{eqnarray}
&&\delta W_{\rm 1\,loop}[u(A_{\rm cl})] \label{traceexp} \\
&&=\sum_{j=1} c_j \int_0^{\infty} d\lambda
\,e^{-\lambda}\,\int d^2x\,\int \frac{d^2k}{(2\pi)^2}
\mbox{tr} \left\{ e^{-ik.x}\,g^{-1} \eta \, e^{\dis \lambda
T^{-1}\Gamma^{0''}/M_j^2}\,g\, e^{+ik.x}\right\}\nonumber\\
&&=\sum_{j=1} c_j \int_0^{\infty} d\lambda
\,e^{-\lambda}\,\int d^2x\int \frac{d^2k}{(2\pi)^2}
\mbox{tr} \left\{ e^{-ik.x}\,\eta_g \, e^{\dis \lambda
g^{-1}T^{-1}\Gamma^{0''} g/M_j^2}\, e^{+ik.x}\right\} \nonumber
\end{eqnarray}
with
\begin{eqnarray}
g^{-1}T^{-1}\Gamma^{0''} g &=& 4\left\{\partial\bar\partial
+2A_{{\rm cl}\,g}\partial+\left( \bar\partial ( A_{{\rm cl}\,g} )
+A_{{\rm cl}\,g}^2 \right) \frac{\partial}{\bar\partial}\right\} \nonu
B_g &=& g^{-1}B g \qq\qq (B=\h, A_{\cl}).
\label{diffop}
\end{eqnarray}
To evaluate eq.(\ref{traceexp}), we split the operator (\ref{diffop})
into two pieces, {\it viz.}  the second derivative $\partial
\bar \partial$ and the rest, called $Y$ from now on. We follow the method
used in \cite{bouwknieuw} for differential operators, but extend here its
use to a non-local differential operator. Applying the
Baker-Campbell-Hausdorff formula,
and rescaling the momenta, we find the following intermediate result
\begin{eqnarray*}
&&\delta W_{\rm 1\,loop}[u(A_{\rm cl})] \\
&&=\sum_{j=1} c_j \int_0^{\infty} d\lambda
\,e^{-\lambda}\, \int d^2x\int d^2p_j\frac{M_j^2}{\lambda (2\pi)^2}
\,e^{-|p_j|^2} \\
&& \hskip 5.mm \mbox{tr} \left\{ \eta_g  e^{-iM_j\,p_j.x/\sqrt\lambda}\,
e^{ \left( 4\lambda Y/M_j^2
+8\lambda^2 \left[ \partial\bar\partial ,Y\right] /M_j^4 +\cdots \right) }
e^{+iM_j\,p_j.x/\sqrt\lambda}\right\}.
\end{eqnarray*}
The next step is to expand the exponential consisting of the non-local
differential operators and let it act on the plane wave.
In the limit $M_j^2\rightarrow\infty$, very few terms will
contribute, but it is not immediately clear how to handle the non-locality.
We have verified that
the so-called pseudo-differential operator calculus (for a general
reference, see \cite{PDO}) works well.
The Leibniz rule is still valid, when written in the form
\begin{eqnarray}
\partial^n\left( f\,*\,\right)= \sum_{k=0}^{\infty}
\frac{n...(n-k+1)}{k!}\left(\partial^k f\right)  \partial^{n-k} *\,.
\label{PDO}
\end{eqnarray}
If $n$ is positive the infinite sum just runs up to $k=n$.
For $n$ negative however, there is an infinite number of terms. Whereas
this might complicate matters, for the
present purpose the series is effectively finite, due to the angular
integral, combined with the $M_j \rightarrow \infty$ limit.
After the angular integral only two pieces remain. In order $M^2_j$,
\begin{eqnarray*}
&&\delta_{\,1} \,W_{\rm 1\,loop}[u(A_{\rm cl})] \\
&&= \sum_{j=1} c_j \int_0^{\infty} d\lambda
\,e^{-\lambda}\, \int d^2x\int d^2p_j\frac{M_j^2}{\lambda (2\pi)^2}
\,\mbox{tr} \left\{ \eta \right\} \,e^{-|p_j|^2} \,,
\end{eqnarray*}
which vanishes since $f^a_{\,\,ab}=0$. As a result the variation of the
effective action contains no infinite terms.
In order $M^0_j$,
\begin{eqnarray*}
&&\delta_{\,\left[\partial\bar\partial ,Y\right] } \,W_{\rm
1\,loop}[u(A_{\rm cl})] \\
&&=\sum_{j=1} c_j \int_0^{\infty} d\lambda
\,e^{-\lambda}\, \int d^2x\int d^2p_j\frac{2\lambda}{\pi^2M_j^2}
\,e^{-|p_j|^2}\\
&& \hskip 10.mm \mbox{tr} \left\{ \eta_g \, e^{-iM_j\,p_j.x/\sqrt\lambda}\,
\left[\partial\bar\partial ,Y \right]
e^{+iM_j\,p_j.x/\sqrt\lambda}\right\}\\
&&=\frac{-1}{\pi}\sum_{j=1} c_j \int d^2x \,\mbox{tr} \left\{ \eta_g \,
\partial (
A_{{\rm cl}\,g}) \right\} \,\int d\left( |p_j|^2\right) \,e^{-|p_j|^2}
\,|p_j|^2 \\
&&=\frac{1}{\pi} \int d^2x \,\mbox{tr} \left\{ \eta \,D[u/k] A_{\rm
cl}\right\} \,. \end{eqnarray*}
Rewriting it in a representation independent way  this becomes
\begin{eqnarray}
\delta W_{\rm 1\,loop}[u(A_{\rm cl})] &=& \frac{-2\tilde h}{2\pi x}\int
d^2x\, \mbox{tr} \left\{ A_{\rm cl}\,\delta \left( u/k \right) \right\}
\nonu
&=& 2\tilde h \int d^2x\, \frac{\partial W^0[u/k]}{\partial u^a/k}
\delta \left( u^a/k \right).\label{anomlike}
\end{eqnarray}
Since there are no gauge invariant polynomials in terms of $u/k$ alone, we
have proven the following one loop result:
\be
W[u] = (k+2\tilde h)\, W^{\,0} \left[ \frac{u}{k}\right]
\label{oneloop}
\ee

\section{Three derivations revisited}

Having completed our one-loop calculation in a regularised framework, we
now reexamine the more heuristic methods reviewed in section three.

It is clear that result \vgl{oneloop} of our calculation,
agrees with the KPZ-result \vgl{kpzresult}, that was based on invariance
under vector-like transformations. The reason for this agreement is not so
clear however. Indeed, the non-local regularisation we introduced, does
not respect the vector-invariance explicitly. In particular, the
gauge field measure is not invariant by itself. But neither is the ghost
field measure. We have not specified how the ghost integral was
regularised, but this is well known: one obtains the standard result in
\vgl{detexplicit} for the chiral determinant by regularising with a
standard (non-invariant) massterm.

To check the KPZ method in the light of our treatment, one has to check
that the quantum theory preserves the vector-invariance. This amounts to a
check of quantum BRST invariance for \vgl{BRSinvact}. It is well known how
to do this. The BRST anomaly  calculation splits into two distinct pieces. The
ghost part calculation can most easily be read off from the result of the
ghost determinant, and is therefore proportional to $2\tilde h \d
W^0[\Ahb]/\d \Ahb \, D[\Ahb]c$. For the gauge field part we  follow the
lines
of \cite{AnomBV}. One then has to determine the expectation value of the
variation of the PV-mass term, just as in the calculation of section
five. In fact, one is just repeating the steps of that section, with the
gauge transformation parameter $\h$ replaced by the BRST ghost $c$. The
result can be read off from \vgl{anomlike}, and exactly cancels the ghost
contribution. This confirms that our
gauge field loop regularisation, when combined with the usual ghost
treatment, preserves the BRST invariance.

Now we also make the connection with the semiclassical reasoning
of section three, based on the factorisation of determinants.
Although the operator equality \vgl{operator} is on solid enough
ground, the statement about the determinants requires a closer
look at regularisations. In section three, we used the standard values for
the determinants of covariant derivatives, \vgl{detexplicit}.
But from the present point of view, this means that certain regularisations
-in fact those regularisations defined by the simplest possible
massterms- are implicit. On the other hand the determinant of the
non-local operator for the $A$-fields  has been defined in its own right in
the preceding section. We can then {\it check} whether the
determinant formula  of (\ref{quotient}) agrees with these choices or not.
We find that it should be replaced by
\begin{eqnarray}
\mbox{det}_{({\rm R})} \frac{\d^2 \, \Gamma^{\,0} [A_{\rm cl}]}
{\d A_{\rm cl} \, \d A_{\rm cl}} =
\frac{\det_{({\rm S})}D[u/k]}{\det_{({\rm S})}\bar D[A_{\rm cl}]} \;
e^{\dis -\frac{2 \tilde h}{2\pi x}\int \mbox{tr} \{ A_{\rm cl}\, u/k \} }
\,,
\label{quotient2}
\end {eqnarray}
where we explicitly indicated our choice of regularisation for the
non-local $A$ determinant by $({\rm R})$, and the standard choices
for the determinants of the covariant derivatives by $({\rm S})$.

It may be interesting to note that there is an alternative way of
writing the semiclassical determinant in a factorised form that {\it does}
agree with our $A$-measure, {\it viz.}
\begin{eqnarray}
\mbox{det}_{({\rm R})} \frac{\d^2 \, \Gamma^{\,0} [A_{\rm cl}]}
{\d A_{\rm cl} \, \d A_{\rm cl}} =
\frac{ \left( \mbox{det}_{({\rm S})} D[u/k] \right) ^2 }
{\det_{({\rm V})} \left\{ \bar D[A_{\rm cl}]\,D[u/k] \right\} } \,,
\label{quotient3}
\end{eqnarray}
where $(\rm V)$ indicates a vector-invariant regularisation.
The log of the determinant in the denominator is proportional to the
covariant induced action of eq.(\ref{covindaction}) in terms of the
chiral gauge fields
$A_{\rm cl}$ and $u/k$. Since these are related to each other through the
Legendre
transform (\ref{u in A}), that determinant equals unity.
For the non-local determinant itself we conclude that, with the present
definitions,
\be
\left( \mbox{det}_{({\rm R})} \frac{\d^2 \, \Gamma^{\,0} [A_{\rm cl}]}
{\d A_{\rm cl} \, \d A_{\rm cl}} \right)^{\frac{1}{2}} =
\mbox{det}_{({\rm S})} D[u/k] = e^{\dis 2\tilde h W^{\,0} [u/k]} \,.
\ee
It may be noted that the determinant factorisation formula
eq.(\ref{quotient}) has been used quite often to calculate
effective actions
\cite{zamo2,PawMeiss,ssvnc,zfactors,gustav}, almost always however in a
rather formal way. Additional insight may be gained by looking back at the
calculation of section four, when interpreting it in terms of these
determinants. From the formal \vgl{detexplicit}, one would have expected
the quadratic terms of the one loop contribution to the effective action
to vanish. Rather, we found \vgl{quad1loop}. It was also pointed out
after \vgl{tweepunt} that a vanishing result would be obtained if
one would regulate by inserting exponential cutoffs by hand. This explains
why this last method gives a result that is the same as the one based on
the straightforward determinant factorisation formula.

\vspace{2mm}
The third derivation in section three was built on a change of variable
including a Jacobian, using the Polyakov-Wiegmann formula, and
assuming the Haar invariance of the measure.
Since we do not intend to give an explicit prescription here for the
regularisation of the $g$-integral (see however \cite{wzwreg,gnw}),
we can not go into this issue in a direct way.
As an alternative, we propose to use formula (\ref{maat}) to {\it define}
the $g$-integration, by relating the so far unspecified
measure $\cd g$ to the measure $\cd A$ we did construct before.
\footnote{We do not follow \cite{Tseytlin} in effectively taking
the square root of this expression. Such a procedure would lead to a
coupling renormalisation factor $k+\th$ instead of $k+2\th$.}
To emphasize the regularisation dependence, we rewrite it as
\begin{equation}
\cd A _{(\rm R)} = \cd g_{(\rm R)} \; e^{\dis 2\tilde{h}S^+[g]} \,.
\label{maatR}
\end{equation}
We now analyse the transformation properties of the resulting measure.
We proceed by computing the transformation
properties of our  well-defined $A$-measure under
finite gauge transformations. This  will teach us how the group measure
itself behaves under left multiplication.
We introduce the finite gauge transforms
\begin{eqnarray*}
A_h &=& \bar\partial h\,h^{-1} + h\,Ah^{-1}\\
\Ahb_h &=& \partial h\,h^{-1} + h\,\Ahb h^{-1}\,.
\end{eqnarray*}
Using the result (\ref{oneloop}), we find to the one-loop order that
\begin{eqnarray}
&&e^{\dis -(k + 2\tilde h)W^0[\Ahb_h]}
\,\,=\,\,\dis \int \cd A_h\,e^{\dis -k\left(\Gamma ^0[A_h] -
\frac{1}{2\pi x}\int \,\mbox{tr}\{A_h\,\Ahb_h\}\right)}\nonu
&&\quad= \dis \int \cd A \,\frac{\cd A_h}{\cd A}\, e^{\dis
-k\left(\Gamma ^0[A]
- \frac{1}{2\pi x} \int\,\mbox{tr}\{A\Ahb\} + W^0[\Ahb_h]
- W^0[\Ahb]\right)}\,.\label{stap1}
\end{eqnarray}
On the other hand, due to the Polyakov-Wiegmann identity, we have that
\begin{eqnarray}
&&e^{\dis -2\tilde h\,W^0[\Ahb_h]}
= e^{\dis -2\tilde h
\left(W^0[\Ahb] + S^- [h] - \frac{1}{2\pi x}\int
\,\mbox{tr}\{h^{-1}\bar\partial h\,\Ahb\}\right)}\nonu
&&= \dis \int \cd A\,e^{\dis -k\left(\Gamma ^0[A] -
\frac{1}{2\pi
x} \int \,\mbox{tr}\{(A + \frac{2\tilde h}{k}h^{-1}\bar\partial
h)\Ahb\} - W^0[\Ahb]\right)
- 2\tilde h S^-[h]}\,.\label{stap2}
\end{eqnarray}
Combining eqs. (\ref{stap1}) and (\ref{stap2}) we find
\begin{eqnarray*}
&&\int \cd A\,e^{\dis -k\left(\Gamma ^0[A] - \frac{1}{2\pi x} \int
\,\mbox{tr}\{A\Ahb\}\right)} \,\frac{\cd A_h}{\cd A} \\
&&= \int \cd A\,e^{\dis -k\left(\Gamma ^0[A - \frac{2\tilde h}{k}
h^{-1} \bar\partial h] - \frac{1}{2\pi x}\int tr\{A\hat{\bar
A}\}\right) - 2\tilde hS^-(h)}
\end{eqnarray*}
where we performed a shift on the quantum field in the second line, such
that in both cases the source $\Ahb$ couples just to $A \equiv
\bar\partial gg^{-1}$.  This enables us to write down the following
Jacobian
\be
\frac{\cd A_{h(\rm R)}}{\cd A_{(\rm R)}} = e^{\dis -2\tilde h\left(S^-[h]
- \int\frac{\delta \Gamma ^0[A]}{\delta A^a}(h^{-1} \bar\partial
h)^a\right)}\,\,.
\label{jac}
\ee
We want to interpret  this $k$-independent factor
as a regulated Jacobian for the change of variables from $A$ to $A_h$. For
consistency, it must obey the group structure for subsequent
transformations; this can be checked by rewriting
\begin{eqnarray}
\frac{\cd A_{h(\rm R)}}{\cd A_{(\rm R)}}
&=& e^{\dis -2\tilde h\left(S^-[h] - \frac{1}{2\pi x} \int
\,\mbox{tr}\{\partial gg^{-1}\, h^{-1}\,\bar\partial h\}\right)}\nonumber\\
&=& e^{\dis -2\tilde h\left(S^-[hg]-S^-[g]\right)}\,,
\label{jacons}
\end{eqnarray}
Using \vgl{maatR} this reduces to the following transformation property for
the group integration:
\[\frac{\cd (hg)_{(\rm R)}}{\cd g_{(\rm R)}} = e^{\dis +M_1[hg] -
M_1[g]} \,,\] with
\begin{eqnarray*}
M_1[g] &=& -2\tilde h\left(S^+[g] + S^-[g]\right)\\
&=& \frac{2\tilde h}{2\pi x}
\int d^2z\,\mbox{tr}\{\bar\partial gg^{-1}\,\partial gg^{-1}\}\,.
\end{eqnarray*}
Our measure $\cd g_{(\rm R)}$ is clearly not invariant,
\footnote{The regularisation dependence of the Haar invariance of the
measure was previously stressed by P. van Nieuwenhuizen.}
but it differs
from an invariant one merely by a local counterterm.  Within the framework
of the WZW model, it is then very natural to incorporate this counterterm
$M_1[g]$ into the action precisely to restore Haar-invariance:
\be
\cd g_{(\rm Haar)}=\cd g_{(\rm R)} \exp -M_1[g].
\ee

We now propose to carry over this counterterm to the non-local theory as
well. The resulting measure will carry the subscript (Haar), the one used
until now will occasionally get the subscript (R).  Although, when
expressed in terms of the field $A$
the counterterm $M_1$ will  of course be non-local, it originates from a
perfectly local expression in $g$, which seems to be an excellent criterion
to characterise the admissible counterterms in the induced WZW model.
We can then recalculate the effective action $W_{\rm Haar}[u]$ that
one obtains when including this counterterm in the action:
\begin{eqnarray*}
&&e^{\dis -W_{\rm Haar}[u]} =\int \cd A_{(\rm Haar)}\,e^{\dis
-k\left(\Gamma
^0[A] - \frac{1}{2\pi x} \int \,\mbox{tr}\{A\,u/k\}\right) }\\
&&\qq= \int \cd A_{(\rm R)}\,e^{\dis -k\left(\Gamma ^0[A] -
\frac{1}{2\pi x} \int \,\mbox{tr}\{A\,u/k\}\right) - M_1[g(A)]}\\
&&\qq= \int  \cd A\,e^{\dis -(k + 2\tilde h)\left(\Gamma ^0[A] -
\frac{1}{2\pi
x} \int \,\mbox{tr}\left\{ A\, \frac{u}{k + 2\tilde h} \right\} \right) +
2\tilde hS^-[g(A)]}\,.
\end{eqnarray*}
Parametrising $\frac{u}{k + 2\tilde h}$ as $\partial\bar
g\bar g^{-1}$ and using the Polyakov-Wiegmann identity, this becomes
\begin{eqnarray*}
&&= e^{\dis -(k + 2\tilde h)W^0\left[ \frac{u}{k
+ 2\tilde h}\right] } \int \cd A\,e^{\dis -(k + 2\tilde h)\Gamma
^0[A_{\bar g^{-1}}] + 2\tilde hS^-[g(A)]}\nonumber\\
&&= e^{\dis -(k + 2\tilde h)W^0\left[ \frac{u}{k + 2\tilde
h}\right]} \int \cd A_{\bar g^{-1}}\,e^{\dis -k\Gamma ^0[A_{\bar g^{-1}}] -
M_1[A_{\bar g^{-1}}]}
\end{eqnarray*}
where the last path integral is just a constant. As a result we have
obtained agreement with formula (\ref{reshaar}) of section three.

\section{Beyond One Loop}
The previous sections only treated the one loop contributions. In the
present section, we extend the analysis to an arbitrary number of loops. We
reconsider the analysis of KPZ, starting from the covariant action
\be
\G [A]+\bar{\G}[\bar A]-\frac{k}{2\pi x}\int
\mbox{tr}\left\{A\,\bar A\right\} \,.
\ee
Fixing the gauge as in section three, and performing the ghost integral,
leads to a fixed value for $\bar A$ and a renormalisation of the $\bar \G$
term, and we have to consider only the $A$-loops.

The theory may be considered to be an approximation to one where the matter
system is strongly coupled to the gauge field. Apart from the induced
action, one then has also a term describing the gauge field itself. Taking
this to be the ordinary Yang-Mills action, it is immediately clear that the
propagators for the gauge field acquire much better convergence properties.
{}From the point of view of the induced action, the Yang-Mills term
constitutes a higher order derivative term. This is reminiscent of the
method used in \cite{FadSlav} for ordinary gauge theories. Since we want to
keep
the vector gauge invariance, the fact that it is covariant is also most
welcome. Let us analyse the divergences in the remaining theory.
The action now is
\[S[A,\Ahb ] = \Gamma ^0[A] - \frac{1}{2\pi x}\int
d^2x\,\mbox{tr}\left\{A\Ahb  + \frac{1}{4\Lambda ^2}F^{\mu \nu
}F_{\mu \nu }\right\}\,,\]
where $\Lambda ^2$ is supposedly a large parameter, that will be sent to
infinity in the end. The Yang-Mills term generates new three and four point
vertices. Explicitly,
\begin{eqnarray*}
\frac{1}{4\Lambda ^2} tr\{F^{\mu \nu }F_{\mu \nu }\} &=& \frac{-2}{\Lambda
^2}\,\mbox{tr}\left\{(\partial A)^2 + (\bar{\partial }\Ahb )^2 +
2\Ahb  \partial \bar{\partial }A \right. \\
&&\left. + 2[A,\Ahb ](\partial A - \bar{\partial }\Ahb
+ \half [A,\Ahb ])\right\}\,.
\end{eqnarray*}
Expansion of the action around the same classical point $A_{\rm cl}$ as
before (see \vgl{u in A} with $u=k \Ahb$) yields
\begin{eqnarray*}
S[A_{\rm cl}+\alpha ,\Ahb ] &=& W^0[\Ahb ] +\half \int
\a^a \frac{\d^2 \, \Gamma^{\,0} [A_{\rm cl}]}
{\d A_{\rm cl}^{\,a} \, \d A_{\rm cl}^{\,b}} \, \a^b+f(A_{\rm cl})\,
\alpha^3+ ... \\
&&-\frac{1}{\Lambda^2\pi x}\int \mbox{tr}\left\{\alpha\, \partial^2\alpha
+ [\alpha,\Ahb ]\left( [\Ahb ,\alpha]
-2\partial\alpha\right) \right\}
\end{eqnarray*}
where we used that $F^{\mu \nu }[A_{\rm cl},\Ahb ] = 0$.  The
kinetic term for the quantum field $\alpha $ is
\[\frac{1}{4\pi x}\int \,\mbox{tr}\left\{\alpha \frac{\partial }
{\bar\partial }\left(1 - \frac{4 \partial\bar\partial}{\Lambda^2} \right)
\alpha \right\}\,,\]
leading to the improved UV behaviour of the Feynman diagrams.
In fact the inclusion of the Yang-Mills term regularises {\it all} the
diagrams made up with the original
$\Gamma ^0[A_{\rm cl}]$-vertices, as can be seen by calculating their
superficial
degree of divergence ($SDD$).  Each vertex adds to the integrand 2
powers of momenta minus the number of internal lines it couples to, or
better. With $V$ the number of vertices and $I$ the number of internal
lines, the total power contributed by the vertices is
$2V - 2I$ or less.  The number of loops $L$ being equal to $I - V + 1$, we
have that $SDD \leq 2-2I=4 - 2L-2V\,$.
The tadpole or (higher order) seagull diagrams are in fact more
convergent than this, because their vertex behaves as an inverse momentum.
Inclusion of the extra Yang-Mills vertices modifies the
formula to $SDD\leq 4-2L-2V_{\rm old}-E_{\rm new}$ where $V_{\rm old}$ is
the number of old vertices, and $E_{\rm new}$ is the number of external
lines coupling to the new vertices.
There are only 2 ill-defined diagrams left in our theory. They are in fact
topologically the same as in fig. 1 and 2, but now with only the new
Yang-Mills vertices present. Since these are one-loop diagrams, we can
regulate them them by adding Pauli-Villars fields with the action
\begin{eqnarray}
S_{PV}[\alpha_i,A_{\rm cl},\Ahb ] &=& \half \int \a_i^{\,a}
\frac{\d^2 \, \Gamma^{\,0} [A_{\rm cl}]}
{\d A_{\rm cl}^{\,a} \, \d A_{\rm cl}^{\,b}} \, \a_i^{\,b}\nonumber\\
&&-\frac{1}{\Lambda^2\pi x}\int \mbox{tr}\left\{\a_i\, \partial^2\a_i
+ [\a_i,\Ahb ]\left( [\Ahb ,\a_i]
-2\partial\a_i \right) \right\}\nonumber\\
&&-\frac{M_i^2}{16\pi x}\int \,\mbox{tr}\left\{\a_i \frac{1}
{\bar{\partial}^2 }\left(1 - \frac{4\partial \bar{\partial
}}{\Lambda^2}\right)\a_i \right\}\,.
\label{pvactie}
\end{eqnarray}
This PV-action has some unfamiliar features. First,
it may be noted that we have chosen a somewhat complicated looking mass
term. This just serves to simplify the resulting Feynman diagrams
(the PV propagators factorise easily with this choice: the denominator
becomes $(|q|^2+\Lambda^2) (|q|^2+M_i^2)$), but is in
fact immaterial: we have checked that all conclusions to be drawn below are
valid also when one leaves out the $M^2/\L^2$ terms. For
$\L\rightarrow\infty$ the mass term is of course the same as \vgl{mass}.
The second point is that the PV-terms only serve to regulate diagrams with
external $\Ahb$ lines, not the propagating $\a$ lines. All loops with
attached $\a$ lines were already made finite by the inclusion of the
higher derivatives. For simplicity, we have therefore decoupled the PV
fields from the quantum $\a$ fields. The coupling to the $A_{\cl}$ is kept
for covariance. Presumably coupling PV-fields according to the
general procedure of duplicating {\it all} couplings gives the same result,
but
we have not checked this explicitly. The decoupling has as a consequence
that the PV-fields do not contribute to higher loop diagrams at all,
not even to compensate divergent subgraphs, since according to the
counting above there should not be any after the introduction of the
Yang-Mills term: only a few diagrams with external
$\Ahb$ and $A_{\cl}$ lines (which do not propagate) are involved in the
additional PV-regularisation.

Having modified the regularisation method for the purpose of rendering
finite the more loop diagrams, it is important to check that the one loop
results of sections four and five remain valid. A priori, they correspond
to the present setup, where the $\L\rightarrow\infty$ limit is taken first.
However, it is not clear whether the order of limits matters. We
will therefore recompute
the 1-loop two point function, {\it i.e.} the quadratic part of
$W[k\Ahb (A_{\rm cl})]$.

The building blocks for our Feynman diagrams are
\begin{eqnarray*}
&&\Delta^i_{ab}(q,r) = g_{\,ab}\,\delta^2(q+r) \,\frac{(|q|^2+\Lambda^2)
(|q|^2+M_i^2)}{q^2 \Lambda^2} \\
&&\sum_{n=3}^{\infty } V^n_{ab}[A_{\rm cl}](q,r) \\
&&\bar V^3_{ab}[\Ahb ](q,r) = \frac{2i}{\Lambda^2(2\pi)^2} \int
d^2p \,\,f_{abc} \,\Ahb ^{\,c}(p) \left( \bar q-\bar r \right) \,\,
\delta^2(p+q+r) \\
&&\bar V^4_{ab}[\Ahb ](q,r) = \frac{-4}{\Lambda^2(2\pi)^4} \int
d^2p \,\,d^2s \,\,f_{ad}^{\,\,\,\,\,c}\,f_{ceb} \,\Ahb^d(p)
\,\Ahb^e(s) \,\,\delta^2(p+q+r+s) \,.\\
\end{eqnarray*}
Note that partly they are expressed in terms of $A_{\cl}$ and partly in
terms of $\Ahb$. For the present calculation it suffices
to put $\Ahb (p) = \frac{\bar{p}}{p}A_{\rm cl}(p)$, and one
has to evaluate the following expression, consisting of 5 different diagrams
(topologically, figs.~1 and 2, but with both the old non-local vertices
and the new Yang-Mills vertices):
\begin{eqnarray*}
&&\frac{1}{2}\sum_i c_i\,\mbox{tr}\left\{\Delta _i^{-1}(V^4 +
\bar{V}^4)\right\} - \frac{1}{4} \sum_i c_i\,\mbox{tr}\left\{\Delta
_i^{-1}(V^3 + \bar{V}^3)\Delta _i^{-1}(V^3 + \bar{V}^3)\right\}\\
&&= -\frac{\tilde h}{(2\pi )^4}\int d^2p\,A_{\rm cl}^{\,a}(p)\left( I_4
+ I_{\bar{4}} + I_{33} + 2I_{3\bar{3}} +
I_{\bar{3}\bar{3}}\right) A_{{\rm cl}\,a}(-p)\,,
\end{eqnarray*}
where the notation reflects the vertices used, the overline denoting the
Yang-Mills vertices.  Let us start with the two diagrams where only the
Yang-Mills vertices are present. The seagull graph vanishes
\be
I_{\bar{4}} = -\frac{2\bar{p}^2}{p^2}\sum_i c_i \int d^2q
\frac{q^2}{(|q|^2 + \Lambda ^2)(|q|^2 + M_i^2)} = 0  \label{i4bar}
\ee
due to the angular integral. The vacuum graph reads
\be
 I_{\bar{3}\bar{3}} =\frac{\bar{p}^2}{p^2}\sum_i c_i \int
d^2q \frac{q^2(p+q)^2(2\bar{q} + \bar{p})^2}{(|q|^2 + \Lambda
^2)(|q|^2 + M_i^2)(|p+q|^2 + \Lambda ^2)(|p+q|^2 +
M_i^2)}.\label{i3bar3bar}
\ee
The numerator is of order $q^6$, to start with, but the PV-contributions
take care of the regularisation. Introducing Feynman parameters,
shifting the momenta, and doing the angular integral, leads to
\begin{eqnarray}
I_{\bar{3}\bar{3}} &=& \sum_i c_i \frac{1}{(M_i^2 - \Lambda ^2)^2}
\int_0^1 d\alpha \int_{\Lambda ^2}^{M_i^2} dx \int_{\Lambda ^2}^{M_i^2}
dy \nonu
&&\hskip 10.mm \int_0^\infty d(|q|^2) \frac{|q|^4A + |q|^2 B + C}{\{|q|^2 +
(1-\alpha )x + \alpha y + \alpha (1-\alpha )|p|^2\}^4} \nonu
\end{eqnarray}
where $A$, $B$ and $C$ depend only on
$p,\bar{p}$ and $\alpha $.  The numerator is now
quartic in the momenta.  It is then clear that $I_{\bar 3\bar{3}}$
will become zero in the limit where $M_i^2$ and $\Lambda ^2$ go to
infinity. Thus, the extra diagrams with only Yang-mills couplings, which
are the only ones having a divergence left before the introduction of
PV-fields, give no contribution. This might have been guessed already
from eqs. (\ref{i4bar}) and (\ref{i3bar3bar}) without introducing the
compensating PV-contributions. However, also in view of \vgl{tweepunt},
we resisted the temptation to apply formal simplifications to unregularised
expressions.

We now proceed with the two ``original'' diagrams.
$$ \begin{array}{l}
\dis I_4 + I_{33} \\
\dis = \sum_i c_i \int d^2q \frac{\Lambda ^2}{p^2}
\frac{(\bar{p} + \bar{q})(p - q)(\bar{p}q -
\bar{q}p)}{(|q|^2 + \Lambda ^2)(|q|^2 + M_i^2)|p + q|^2} \\
\dis \hskip 2.mm + \int d^2q \frac{\Lambda ^4}{p^2} \frac{(\bar{p}q -
\bar{q}p)^2}{(|q|^2 + \Lambda ^2)(|q|^2 + M_i^2)(|p + q|^2 + \Lambda
^2)(|p + q|^2 + M_i^2)} \\
\dis = \sum_i c_i \frac{2\pi \Lambda^2}{(M_i^2 - \Lambda ^2)}
\frac{|p|^2}{p^2} \int_0^1 d\alpha \int_{\Lambda ^2}^{M_i^2} dx \frac{(1 -
\alpha ) }{\{(1 - \alpha )x + \alpha (1 - \alpha )|p|^2\}} \\
\dis \hskip 2.mm - \frac{2\pi \Lambda ^4}{(M_i^2 - \Lambda ^2)^2}
\frac{|p|^2}{p^2} \int_0^1 d\alpha \int_{\Lambda ^2}^{M_i^2} dx
\int_{\Lambda ^2}^{M_i^2} dy \frac{\alpha (1 - \alpha )}{\{ (1
- \alpha )x + \alpha y + \alpha (1 - \alpha )|p|^2\}^2}
\end{array} $$
Here it matters in which order the regulators are removed.  If the
Pauli-Villars masses go to infinity, $\Lambda ^2$ remaining finite, the
Pauli-Villars terms drop out, and
\begin{eqnarray*}
\lim_{\L\rightarrow\infty}\lim_{M_i\rightarrow\infty}(I_4 + I_{33}) &=&
\frac{-2\pi |p|^2}{p^2}\lim \int_0^1 d\alpha \ln \frac{\alpha
\Lambda ^2 + \alpha (1 - \alpha )|p|^2}{\Lambda ^2 + \alpha (1 - \alpha
)|p|^2}\\
&=& \frac{2\pi |p|^2}{p^2}\,.
\end{eqnarray*}
In the opposite order one finds that
\begin{eqnarray*}
\lim_{M_i\rightarrow\infty}\lim_{\L\rightarrow\infty}(I_4 + I_{33})
 &=&\lim\sum_i c_i \frac{-2\pi |p|^2}{p^2} \int_0^1 d\alpha  \ln
\frac{\alpha M_i^2 + \alpha (1 - \alpha )|p|^2}{M_i^2 + \alpha (1 - \alpha
)|p|^2 }\\
&=& \frac{-2\pi |p|^2}{p^2}\,,
\end{eqnarray*}
which is of course the result \vgl{result1} that we obtained in
section four without introducing $\L$ in the first  place.

The remaining contribution equals
$$\begin{array}{l}
\dis 2I_{3\bar{3}} \\
\dis =\sum_i c_i \int d^2q \frac{2\Lambda
^2}{p^2} \frac{(\bar{p}q
- \bar{q}p)\,q\,\bar{p}\,(2\bar{q} + \bar{p})\,(p + q)}{(|q|^2
+ \Lambda ^2)(|q|^2 + M_i^2)(|p + q|^2 + \Lambda ^2)(|p +
q|^2 + M_i^2)}\\
\dis = \sum_i c_i \frac{-8\pi \Lambda ^2}{(M_i^2 - \Lambda ^2)^2}
\frac{|p|^2}{p^2} \int_0^1 d\alpha  \int_{\Lambda ^2}^{M_i^2} dx
\int_{\Lambda ^2}^{M_i^2} dy \frac{\a (1-\a )}{\{(1-\a )x +
\a y + \a (1-\a )|p|^2\}}\\
\dis \hskip 3.mm - \frac{4\pi \Lambda ^2}{(M_i^2 - \Lambda ^2)^2}
\frac{|p|^4}{p^2}
\int_0^1 d\alpha  \int_{\Lambda ^2}^{M_i^2} dx \int_{\Lambda ^2}^{M_i^2} dy
\frac{\alpha ^3(1 - \alpha )}{\{(1 - \alpha )x + \alpha y + \alpha (1 -
\alpha )|p|^2\}^2}\,.
\end{array} $$
The second term is equal to zero, no matter how the limits are taken.
The rest gives
\begin{eqnarray*}
\lim_{\L\rightarrow\infty}\lim_{M_i\rightarrow\infty}2I_{3\bar{3}} &=&
\lim -\frac{8\pi |p|^2}{p^2} \int_0^1 d\alpha
\ln \Lambda ^2 - \alpha\ln(\alpha \Lambda ^2) -(1-\a ) \ln((1-\alpha )
\Lambda ^2) \\
&=& -\frac{4\pi |p|^2}{p^2}
\end{eqnarray*}
and
\begin{eqnarray*}
\lim_{M_i\rightarrow\infty}\lim_{\L\rightarrow\infty}2I_{3\bar{3}} &=&
\frac{16\pi |p|^2}{p^2} \sum_i c_i \int_0^1
d\alpha \,\alpha \ln \alpha \\
&=& 0\,.
\end{eqnarray*}
Summing all five diagrams we have, for both limits,
\[W_{\rm quantum}[k\Ahb (A_{\rm cl})] = -\frac{\tilde h}{2\pi x} \int
d^2x\,\mbox{tr}\left\{A_{\rm cl} \frac{\partial }{\bar{\partial} } A_{\rm
cl} + \cdots \right\}. \]
We have checked that all other $\L,M_i\rightarrow\infty$ limits give the
same answer too.
It is satisfying that the sum of all one loop diagrams
{\it does not} depend on the specific way of taking the limits,
despite the fact that expressions for individual diagrams do.
We conclude that the one loop results of section four remain valid.

To extend these result to all loops, we now continue along the line of
section five, and compute  the full quantum effective action through its
(anomalous) gauge variation. Again, using the standard ghost determinant,
we only have to calculate the gauge field contribution, with the aim to
show that it cancels the ghost anomaly.

Since the regulating Yang-Mills term is invariant under the simultaneous
transformation of classical and quantum fields, the Pauli-Villars
mass term is still the only symmetry breaking part of the Lagrangian. The
variation of $W_{\rm quantum}[k\Ahb ]$ can thus be written as the
expectationvalue of the variation of the mass term of eq.(\ref{pvactie}).
When expanding this expectationvalue in a series of Feynman diagrams,
it can be seen that the only contributing diagrams are those with
one loop of P.V. fields  (and no other loops).

The full variation of $W_{\rm quantum}[k\Ahb ]$ is therefore given by
\begin{eqnarray}
\delta W_{\rm quantum}[k\Ahb ] = \half \sum_{j=1} c_j M_j^2\,
\mbox{Tr} \left\{ X \,\left(
\Delta^j + \sum_{n=3} \left( V^n [A_{\rm cl}(\Ahb )] + \bar V^n[\Ahb
]\right) \right)^{-1} \right\}\, ,\label{full}
\end{eqnarray}
where $X$ results from varying the mass term:
\be
X_{ab}(q,r) = \frac{1}{(2\pi)^2} \int
d^2p \,\,f_{abc} \,\eta^{\,c}(p) \left\{  \frac{1}{r^2} \left(
\frac{\Lambda^2 + |r|^2}{\Lambda^2} \right) -\frac{1}{q^2}\left(
\frac{\Lambda^2 + |q|^2}{\Lambda^2} \right) \right\} \,\, \delta^2(p+q+r)
\,.
\ee

The term without vertex vanishes because $tr\{ \eta \} $ is $0$ .
The next term is
\begin{eqnarray*}
\delta_{\bar 3 }\, W_{\rm quantum}[k\Ahb ] &=& -\half \sum_{j=1} c_j
M_j^2\,
\mbox{Tr} \left\{ X \, \Delta ^{-1}_j \,\bar V^3[\Ahb ] \,\Delta^{-1}_j
\right\}\\
&=& -\frac{\tilde h}{(2\pi )^4}\int d^2p\,\Ahb ^{\,a}(p)\, J_{\bar{3}}\,
\eta_a (-p)
\end{eqnarray*}
with
\begin{eqnarray}
J_{\bar{3}}
&=& \sum_{j=1} c_j M^2_j \int d^2q
\frac{2\,i \,q^2 \,(\bar p+2\bar q)}{(|q|^2 + \Lambda ^2)(|q|^2 + M_j^2)
(|p+q|^2 + M_j^2) } \nonu
&=& \sum_{j=1} c_j \frac{-8\pi i p \,M_j^2}{(M_j^2 - \Lambda ^2)}
\int_0^1 d\alpha \int_{\Lambda ^2}^{M_j^2} dx
\frac{ \a (1 - \alpha ) }{\{(1 - \alpha )x + \a M_j^2 +\alpha (1 - \alpha
)|p|^2\}}.       \label{J3bar}
\end{eqnarray}
In the beginning of this section we demonstrated that the order of limits
was immaterial for the sum of all contributions. Assuming that this extends
to the present case, we evaluate
\begin{eqnarray}
\lim_{\L\rightarrow\infty}\lim_{M_j \rightarrow \infty }
J_{\bar 3} &=& -8\pi i p\,\lim\sum_{j=1} c_j   \int_0^1 d\alpha \,\a
\ln
\frac{M_j^2 + \alpha (1 - \alpha )|p|^2}{\a M_j^2 + (1-\a )\Lambda^2  +
\alpha (1 - \alpha )|p|^2 }\nonumber\\
&=& 2\pi i p \,,\label{J3barend}
\end{eqnarray}
leading to
\begin{eqnarray}
\delta_{\bar 3 }\, W_{\rm quantum}[k\Ahb ] &=&
\frac{2\tilde h}{2\pi x}\int
d^2x\, \mbox{tr} \left\{ \eta \, \bar\partial \Ahb \right\}  \nonu
  &=& \delta \left( 2\tilde h W^0[\Ahb ] \right) \,. \label{deltaWq}
\end{eqnarray}
Remarkably, with this order of limits, this single diagram already
furnishes
the entire quantum variation we expected to obtain. It remains to show
that all the other graphs vanish.

As an illustration we compute the first term explicitly:
\begin{eqnarray*}
\delta_3\, W_{\rm quantum}[k\Ahb ] &=& -\half \sum_{j=1} c_j
M_j^2\,
\mbox{Tr} \left\{ X \, \Delta ^{-1}_j \,V^3[A_{\rm cl}(\Ahb) ]
\,\Delta^{-1}_j \right\}\\
&=& -\frac{\tilde h}{(2\pi )^4}\int d^2p\,A_{\rm cl}^{\,a}(p)\,
J_3\, \eta_a (-p)
\end{eqnarray*}
with
\begin{eqnarray}
J_3 &=& \sum_{j=1} c_j M^2_j \int d^2q
\frac{\Lambda^2\,i \,(\bar pq-\bar qp)^2}{p|q|^2(|q|^2 +
M_j^2)(|p+q|^2+\Lambda^2) (|p+q|^2 + M_j^2) } \label{J3} \\
&=& \sum_{j=1} c_j \frac{-2\pi i \bar p \,\Lambda^2}{(M_j^2 - \Lambda
^2)} \int_0^1 d\alpha \int_{0}^{M_j^2} dx \int_{\Lambda ^2}^{M_j^2} dy
\frac{ \a (1 - \alpha ) }{\{(1 - \alpha )x + \a y +\alpha (1 - \alpha
)|p|^2\}^2} \nonumber
\end{eqnarray}
In the last line, there is a factor $M^{-2}$ multiplying an integral that
diverges only logarithmically when $M\rightarrow\infty$. This
can already be seen from the first line of \vgl{J3}, by neglecting the
momentum dependence of the two factors $M_j^2+|k|^2$, and counting the
degree of divergence of the remaining $q$-integral.
Consequently,
\be \lim_{\L\rightarrow\infty}\lim_{M_j \rightarrow \infty }
J_{3} = 0.
\ee
The same argument can be repeated for the diagrams with more and/or other
insertions. Again, by moving in front two factors $M^{-2}$ and counting the
degree of divergence of the remaining momentum integral, one finds at most
logarithmic divergences, except in a few cases. Of these, one was
computed in \vgl{J3barend}. The other two have one $\bar V^4$ or two $\bar
V^3$ insertions, and vanish also, by an explicit check. Consequently, with
this order of the limits, \vgl{deltaWq} gives indeed the total
contribution.

If we check what happens when we again reverse the limits, we have
\begin{eqnarray*}
\lim_{M_j \rightarrow\infty}\lim_{\L \rightarrow \infty }
J_{3} &=& -2\pi i \bar p\,\sum_{j=1} c_j   \int_0^1 d\alpha \ln
\frac{M_j^2 + \alpha (1 - \alpha )|p|^2}{\a M_j^2 +
\alpha (1 - \alpha )|p|^2 }\\
&=& 2\pi i \bar p \,,
\end{eqnarray*}
while the $\frac{1}{\Lambda^2}$ of the $\bar V^3[\Ahb ]$ vertex
makes $J_{\bar 3}  = 0 $ in this limit (see \vgl{J3bar}). Using the
relation between $A_{\cl}$ and $\Ahb$, which to the first order is $\bar p
A_{\cl}(p) = p\Ahb(p)$, we recover the
full quantumvariation of eq.(\ref{deltaWq}), plus additional terms with
higher powers of the field $\Ahb$. These additional terms only drop out
when including the contributions to eq.(\ref{full}) of the P.V. diagrams
with more $V^n[A_{\rm cl}]$ vertex insertions. These diagrams indeed give
finite results in the limit we are considering \footnote{In fact
we already calculated these diagrams in section 5,
albeit in configurationspace.}, such that the
complete expression will emerge this time in the form of an infinite series
\be
\delta \left( 2\tilde h W^0[\Ahb (A_{\rm cl}) ] \right) =
\frac{2\tilde h}{2\pi x}\int
d^2x\, \mbox{tr} \left\{ \eta \,D\left[ -2\p\frac{\d \Gamma^{\,0} [A_{\rm
cl}]} {\delta A_{\rm cl}} \right] A_{\rm cl} \right\}
\ee

In order to calculate the full effective action, it thus suffices to study
its 1-loop variation. This 1-loop variation can be obtained in two ways:
either as an infinite
sum of diagrams made up with the 'original' non-local vertices; or
by using a different order of limits, in which case
all the information collapses in a single diagram.

\section{Conclusions}
Especially for two-dimensional theories, there has been a marked increase
in the interest for non-local field theories, mainly when arising as
induced theories describing the dynamics of degrees of freedom that arise
through anomalies. Whereas for local field theories experience has
accumulated considerably over the years, for non-local field theories a lot
remains to be learned. In this paper we have shown that, as far as
regularisation is concerned, one can go a long way with rather conventional
methods. We have shown that Pauli-Villars methods can be extended quite
easily, provided one is willing to introduce non-local mass terms. Also the
method of higher covariant derivatives has proven useful in limiting the
breaking of symmetries to a (presumably unavoidable) core. We have also
shown that anomaly-style calculations have an equally enlarged field of
application, and here also some extension of traditional methods (we used
pseudo-differential operators) was needed. We have proposed a scheme
for a specific class of non-local induced actions, {\it viz.} gauge field
actions induced through coupling with affine currents, although without any
doubt it can be extended to a large class of similar actions. We have used
our scheme to clear up some discrepancies between different calculations of
the (finite) renormalisation factors that allow one to express the
effective quantum actions of these theories in terms of their classical
actions. The conclusion here is, that different counterterms do indeed lead
to different field renormalisation factors, but no inconsistency
arises if one adopts a given regularisation procedure systematically ---
although it may not be obvious how to guarantee this in
practice. We have shown the precise
relation between different formal evaluations, in particular how different
symmetry requirements, vector gauge symmetry conservation and Haar
invariance, are related through a counterterm, that is local in the
framework of a WZW lagrangian. It may be noted here that there exist other
methods where the regularisation is much more implicit, as for example in
point splitting methods combined with operator product expansions.
We have not established a direct connection between that method and the
regularisation(s) in this paper, but expect that there  should be no
objection to it either, if adopted systematically. We remind the reader
that it leads to results 'halfway' between the vector-invariant and the
Haar-invariant results.

One attempt to bring these non-local theories back to the more familiar
local ground would be through a transformation of variables. For example,
for induced gravity the Polyakov variables make the action local. The
resulting local action however is rather involved, and the application of
standard perturbative renormalisation theory is not straightforward. Since
universal normalisation requirements
 for the resulting theories are lacking, we have been guided by the
symmetry preservation of the theory, and have adopted the working rule not
to add any counterterms unless required by symmetry. One can arrive at the
same rule when considering the theory treated in section seven. The
non-local action can be considered as arising from a local matter action,
where one also includes the Yang-Mills term from the start. The
renormalisation conditions and counterterms for this local field theory
then will not involve the non-local vertices of the induced theory either.
One has noticed (see section four) that we never {\it needed} to introduce
counterterms
to cancel infinities. The same property also invites one to perform formal
calculations without even mentioning the divergences, although
we have seen that the finite results obtained in this way may be
incorrect. Physically, it is linked to the {\it induced} nature
of the theories we considered (which may be the only non-local theories
that one wants to consider anyway), which leads to much softer ultraviolet
behaviour than expected generically.

\vspace{10mm}
{\bf Acknowledgements}: We would like to thank Jan de Boer, Jose
Figueroa,  Toine Van Proeyen, A. Tseytlin and especially Peter van
Nieuwenhuizen and Bernard de Wit for discussions.


\begin{thebibliography}{88}
\bibitem{wzw}
J.Wess, B.Zumino, Phys.\ Lett.\ {\bf B37}(1971)95.\\
  E. Witten, \cmp {\bf 92} (1984) 455.
\bibitem{polouche} A. M. Polyakov, in 'Fields, Strings
and Critical Phenomena', eds. E. Br\'ezin and J. Zinn-Justin,
North-Holland (1990) (Les Houches 1988)
\bibitem{kpz} V.G. Knizhnik, A.M. Polyakov and A.B. Zamolodchikov,
  Mod. Phys. Lett. {\bf A3} (1988) 819.
\bibitem{ssvnc}
  K. Schoutens, A. Sevrin and P. van Nieuwenhuizen,
  Nucl. Phys. {\bf B371} (1992) 315.
\bibitem{dbg}
  J. de Boer and J. Goeree, Utrecht preprint THU-92/33\\
  J. de Boer, Doctoraatsthesis, Utrecht 1993.
\bibitem{zfactors}
  A. Sevrin, K. Thielemans and W. Troost,
  LBL-33738, UCB-PTH-93/06, KUL-TF-93/09.\\
  A. Sevrin and W. Troost,
  LBL-34125, UCB-PTH-93/19, KUL-TF-93/21.
\bibitem{drisok} V. G. Drinfeld and V. V. Sokolov, J. Sov. Math. {\bf 30}
 (1984) 1975          \\
L. Feher, L. O'Raifeartaigh, O. Ruelle, I. Tsutsui and
A. Wipf, Phys. Rep. {\bf 222} (1992) 1.
\bibitem{Pauli} W.~Pauli and P.~Villars, Rev.\ Mod.\ Phys.\ {\bf 21}
(1949) 434.
\bibitem{polwi}
  A. M. Polyakov and P. B. Wiegmann, \pl {\bf 131B} (1983) 121;
  \pl {\bf 141B} (1984) 223.
\bibitem{zamo2}
  Al.B. Zamolodchikov, preprint ITEP 84-89 (1989).
\bibitem{stony}
  K. Schoutens, A. Sevrin and P. van Nieuwenhuizen, in the proceedings of the
Stony Brook conference {\it Strings and Symmetries 1991}, (World Scientific,
1992).
\bibitem{orlando}
  O. Alvarez, \np {\bf B238} (1984) 61
\bibitem{loopsnlft}
 M. T. Grisaru and P. van Nieuwenhuizen, Int. Jour. Mod. Phys. {\bf A7}
(1992) 5691.
\bibitem{FrankRuud1} F.~De~Jonghe, R.~Siebelink, W.~Troost, Phys. Lett. {\bf
B288} (1992) 47.
\bibitem{diaz}
 A. Diaz, W. Troost, P. van Nieuwenhuizen and A. Van Proeyen,
 Int. Jour. Mod. Phys. {\bf A4} (1989) 3959.
\bibitem{bouwknieuw}
 P. Bouwknegt and P. van Nieuwenhuizen, Class. Quantum Gravity {\bf 3}
(1986) 207
\bibitem{PDO}
M. A. Shuba, "Pseudodifferential Operators and Spectral Theory", Springer
Verlag 1987.
\bibitem{AnomBV}
 W. Troost, P. van Nieuwenhuizen and A. Van Proeyen, \np {\bf B333} (1990)
727.
\bibitem{PawMeiss} K.A. Meissner and J. Pawe\sss czyk, Mod. Phys. Lett. {\bf
A5} (1990) 763.
\bibitem{gustav}
   G.W. Delius, M.T. Grisaru and P. van Nieuwenhuizen, preprint
   CERN-TH.6458/92.
\bibitem{wzwreg}
 M. Bos, \pl {\bf B189} (1987) 435 \\
 H. Leutwyler and M. Shifman, Int. Jour. Mod. Phys. {\bf A7} (1992) 795
\bibitem{gnw}
 M. T. Grisaru, B. de Wit and P. van Nieuwenhuizen, Brandeis-Stony
Brook-Utrecht preprint.
\bibitem{Tseytlin}
 A. Tseytlin, Imperial college preprint 12/92 and CERN-TH.6804/93.
\bibitem{FadSlav}
 L. D. Faddeev and A. A. Slavnov, {\it Gauge Fields: Introduction
to Quantum Theory}, Frontiers in Physics series 50, The Benjamin/Cummings
Publishing Company, Reading, Massachussetts (1980).
\end{thebibliography}
\end{document}